\documentclass[11pt]{article}

\usepackage[utf8]{inputenc}
\usepackage{amsmath, amssymb, amsthm}
\usepackage[letterpaper, margin=1in]{geometry}
\usepackage{titlesec}
\usepackage{algorithm}      
\usepackage{algpseudocode}

\usepackage{booktabs}
\usepackage{graphicx}
\usepackage{caption}
\usepackage{microtype}
\usepackage{enumitem}
\usepackage[table]{xcolor}
\usepackage{float}

\usepackage{setspace}
\usepackage[authoryear, round]{natbib} 

\usepackage[colorlinks=true, citecolor=blue, linkcolor=blue, urlcolor=blue]{hyperref}

\title{\textbf{Objective-Driven Ensembles: Bridging the Gap Between Interpretable Sparsity and Algorithmic Prediction \vspace{0.75cm}}}

\author{
	\Large \textbf{Anthony Christidis$^*$} \\
	Department of Statistics, University of British Columbia \\
	Department of Biomedical Informatics, Harvard Medical School \\
	$^*$Corresponding Author: \href{mailto:anthony-alexander_christidis@hms.harvard.edu}{\texttt{anthony-alexander\_christidis@hms.harvard.edu}} \\[1.5em]
	\Large \textbf{Stefan Van Aelst} \\
	Department of Mathematics, KU Leuven \\[1.5em]
	\Large \textbf{Ruben Zamar} \\
	Department of Statistics, University of British Columbia
}

\date{} % Leave blank for submission

\begin{document}
	
	\maketitle
	
	\begin{abstract}
		Sparse methods (e.g., Best Subset Selection, Elastic Net) are the standard approach for obtaining interpretable models, but they can suffer from high variance and vulnerability to spurious correlations. Alternatively, algorithmic ensembles (e.g., Random Forests, Gradient Boosting) achieve high prediction accuracy but yield uninterpretable black boxes driven by randomization or sequential residual fitting. In recent years, a unifying paradigm has emerged: Objective-Driven Ensembles. By generalizing best subset selection into a joint mathematical optimization problem, this approach generates interpretable ensembles by optimally splitting predictors across a small number of diverse models. In this paper, we synthesize this growing body of literature and illustrate the statistical principles driving its empirical success. Specifically, we utilize finite-sample bounds to demonstrate how penalizing predictor overlap controls ensemble covariance and provides a mathematical hedge against spurious correlations. We evaluate these mechanics using an exact combinatorial oracle, and review how recent computational approximations have successfully scaled this framework to a variety of domains, including high-dimensional data, classification tasks, and settings with casewise or cellwise contamination, achieving machine-learning-level accuracy while retaining the interpretability of sparse models.
	\end{abstract}
	
	\vspace{0.5cm}
	\vspace{0.5cm}
	
	\newpage
	
	% ---------------------------------------------------------
	\section{Introduction}
	\label{sec:intro}
	
	The analysis of high-dimensional data has long been characterized by a fundamental trade-off between interpretability and prediction accuracy. In a seminal paper, \citet{breiman2001two} characterized this divide as the ``Two Cultures'' of statistical modeling. On one side, data modeling emphasizes interpretable, often sparse, parametric structures. Methods such as Best Subset Selection \citep{bertsimas2016best, hastie2020best}, the Lasso \citep{tibshirani1996regression}, and the Elastic Net \citep{zou2005elasticnet} successfully isolate the most important predictors, but they typically exhibit higher variance and can suffer in predictive performance compared to unconstrained models. On the other side, algorithmic modeling techniques, such as Random Forests \citep{breiman2001random} and Gradient Boosting \citep{friedman2001greedy}, combine a large collection of models to achieve exceptional prediction accuracy. However, these ensemble models are driven by randomization or sequential residual fitting, resulting in black-box models that obscure the underlying relationships between the predictors and the response.
	
	A primary reason interpretable sparse models struggle to match the accuracy of algorithmic ensembles is their handling of correlated predictors and spurious noise. In high-dimensional settings, the Lasso tends to arbitrarily select a single variable from a group of correlated predictors, leading to highly unstable models \citep{zhao2006model, buhlmann2011statistics}. The Elastic Net significantly improves upon this limitation via the ``grouping effect,'' which shrinks the coefficients of highly correlated variables together. However, the Elastic Net still fundamentally relies on a single-model architecture. If the selected group of variables contains spurious correlations driven by finite-sample noise or selection bias, the resulting single model becomes highly vulnerable to generalization failure when evaluated on unseen data \citep{fan2016guarding}.
	
	While methods like rule extraction \citep{friedman2008predictive} or Bayesian tree priors \citep{chipman2010bart} attempt to bridge this gap, they still fundamentally rely on algorithmic architectures. To fully reconcile interpretability with predictive power, \emph{Objective-Driven Ensembles} have emerged as a unifying paradigm. Dispensing with randomization and sequential fitting, this approach formulates ensemble generation as a joint mathematical optimization problem. By generalizing Best Subset Selection to simultaneously estimate multiple sparse models while penalizing predictor overlap, it naturally partitions correlated variables across distinct models. Recent advances have successfully realized this objective across various settings. Initial approaches used multi-convex relaxations to approximate the splitting penalty for linear \citep{christidis2020split} and logistic regression \citep{christidis2025data, yang2026diverse}. The framework has since been scaled to exact $\ell_0$-constrained formulations via projected gradient algorithms \citep{christidis2025multi}, alongside robust extensions for both casewise \citep{christidis2026robust} and cellwise contamination \citep{christidis2026fast}. Ultimately, the split ensemble provides a reliable middle ground: averaging predictions improves robustness, while the final model remains highly interpretable as a small, discrete set of sparse equations.
	
	While fast computational algorithms for this splitting objective have proven highly successful in applied tasks, the underlying statistical mechanics driving their performance have not been formally synthesized. In this paper, we unify this growing body of literature into a cohesive statistical paradigm and illustrate the fundamental principles that explain its empirical success.
	
	The remainder of the article is organized as follows. In Section \ref{sec:idealized}, we define the exact combinatorial oracle for the splitting paradigm, which we refer to as Exact Multi-Model Subset Selection (MSS). In Section \ref{sec:theory}, we examine the statistical properties of this objective, utilizing finite-sample bounds to demonstrate how penalizing predictor overlap controls prediction covariance and acts as a mathematical hedge against generalization error driven by spurious sample correlations. Section \ref{sec:exact_empirical} empirically evaluates these mechanics in a controlled, low-dimensional setting, completely free of approximation error. Because the exact combinatorial oracle is NP-hard, Section \ref{sec:review} reviews how recent computational approximations have successfully scaled this framework to high-dimensional, classification, and robust settings. In Section \ref{sec:high_dim}, we benchmark these tractable approximations against state-of-the-art machine learning algorithms on high-dimensional data. Section \ref{sec:discussion} concludes and discusses avenues for future research.
	
	% ---------------------------------------------------------
	\section{Exact Multi-Model Subset Selection (MSS)}
	\label{sec:idealized}
	
	Best Subset Selection (BSS) evaluates all possible combinations of candidate predictors to identify the single model that minimizes a predefined goodness-of-fit criterion. For a response vector $\mathbf{y} \in \mathbb{R}^n$ and a standardized design matrix $\mathbf{X} \in \mathbb{R}^{n \times p}$, exact BSS solves the non-convex optimization problem:
	\begin{equation}
		\label{eq:bss}
		\min_{\boldsymbol{\beta} \in \mathbb{R}^p} \mathcal{L}(\mathbf{y}, \mathbf{X}\boldsymbol{\beta}) \quad \text{subject to} \quad \|\boldsymbol{\beta}\|_0 \le t,
	\end{equation}
	where $\mathcal{L}(\cdot, \cdot)$ is a designated empirical loss function (e.g., squared error for continuous responses) and $t \le \min(p, n-1)$ dictates the maximum model size. To bridge the gap between interpretable sparse modeling and variance-reducing ensemble techniques, Multi-Model Subset Selection (MSS) \citep{christidis2025multi} generalizes the BSS framework. Rather than searching for a single optimal subset, MSS seeks to optimally partition the candidate predictors among a fixed number of $G \ge 2$ sparse models, which are subsequently combined to form an ensemble. 
	
	Exact MSS is formulated as a joint optimization problem that simultaneously estimates the $G$ coefficient vectors $\boldsymbol{\beta}^{1}, \dots, \boldsymbol{\beta}^{G} \in \mathbb{R}^p$. The objective is to minimize the aggregate loss across all models subject to sparsity and diversity constraints:
	\begin{equation}
		\label{eq:exact_mss}
		\min_{\boldsymbol{\beta}^1, \dots, \boldsymbol{\beta}^G \in \mathbb{R}^p} \sum_{g=1}^G \mathcal{L}(\mathbf{y}, \mathbf{X}\boldsymbol{\beta}^g) \quad \text{subject to} \quad 
		\begin{cases} 
			\|\boldsymbol{\beta}^g\|_0 \le t, & 1 \le g \le G, \\
			\sum_{g=1}^G \mathbb{I}\left(\beta_j^g \neq 0\right) \le u, & 1 \le j \le p.
		\end{cases}
	\end{equation}
	In \eqref{eq:exact_mss}, $\mathbb{I}(\cdot)$ denotes the indicator function. The procedure is governed by two tuning parameters, typically selected via cross-validation, where $t$ dictates the $\ell_0$ sparsity of each model, and $u \ge 1$ serves as a diversity budget restricting the maximum number of models that can share a single predictor.
	Explicitly parameterizing this overlap allows the ensemble to adapt to the underlying data. If the signal is driven by a single unambiguous set of strong predictors, a large $u$ allows all models to leverage them; if $u$ is entirely unconstrained, the procedure trivially collapses to estimating the same BSS model $G$ times. Conversely, in the presence of highly correlated proxy variables, restricting $u$ forces the algorithm to explore distinct explanatory pathways, yielding diverse models that maximize variance reduction. Finally, the optimal models are aggregated, typically via an unweighted average of their predictions or fitted probabilities, to form the ensemble.
	
	A defining characteristic of the exact MSS objective in \eqref{eq:exact_mss} is its capacity for global feature selection. Because each of the $G$ models is constrained to a maximum size $t$, the algorithm selects a total subset of $k \le \min(p, G \times t)$ unique informative predictors from the full feature space to distribute across the ensemble. The remaining $p - k$ predictors are assigned coefficients of exactly zero across all $G$ models. Consequently, unlike heuristic ensembles (such as Random Forests) that tend to eventually incorporate uninformative noise variables across their deep collection of trees, the MSS ensemble actively filters out spurious features, resulting in strict global sparsity and high interpretability.
	
	While mathematically elegant, obtaining the global minimum of \eqref{eq:exact_mss} requires an exhaustive search over all valid partitions of the feature space. Let $p_g$ denote the number of active variables allocated to group $g$, and let $q = \sum_{g=1}^G p_g$ be the total number of distributed features. Furthermore, let $h_i(p_1, \dots, p_G)$ denote the number of elements in the sequence $p_1, \dots, p_G$ that are exactly equal to $i$. As derived by \citet{christidis2020split}, the total number of valid configurations to split $p$ features into $G$ strictly disjoint groups ($u=1$) bounded by maximum size $t$ is given by the combinatorial formula:
	\begin{equation}
		\label{eq:combinatorics}
		\mathcal{C}(p, G, t) = \sum_{p_1 \le \dots \le p_G \le t} \binom{p}{q} \left[ \frac{q!}{p_1! \dots p_G!} \prod_{i=1}^t \frac{1}{h_i(p_1, \dots, p_G)!} \right].
	\end{equation}
	This search space grows explosively with the number of candidate predictors. For example, evaluating a highly restricted setting with $p=15$ predictors split into $G=3$ models of maximum size $t=10$ requires computing $\mathcal{C}(15, 3, 10) = 171,761,951$ unique subset configurations. Furthermore, if predictors are permitted to overlap between models ($u > 1$), the combinatorial space expands even more drastically. 
	
	Because evaluating the combinatorics in \eqref{eq:combinatorics} constitutes an NP-hard problem \citep{welch1982algorithmic}, Exact MSS cannot be directly deployed on modern high-dimensional datasets. Instead, it serves as the idealized theoretical oracle of the objective-driven ensemble paradigm. In Section \ref{sec:theory}, we use this exact oracle to illustrate the statistical principles driving the splitting objective. In Section \ref{sec:exact_empirical}, we evaluate this oracle empirically to demonstrate its mechanics free of approximation error. Finally, these insights motivate the development of the computationally tractable algorithms reviewed in Section \ref{sec:review}.
	
	% ---------------------------------------------------------
	\section{Theoretical Properties of Objective-Driven Splitting}
	\label{sec:theory}
	
	While fast computational algorithms for objective-driven ensembles demonstrate strong empirical performance, their theoretical underpinnings have not been formally unified. In this section, we present two primary statistical properties of the Exact MSS framework under squared error loss. Rather than relying on asymptotic assumptions ($n \to \infty$) which often fail to hold in high-dimensional settings, we utilize deterministic, finite-sample bounds to illustrate the mechanics of the splitting objective. Detailed algebraic derivations for these bounds are provided in the Appendices~\ref{sec:AppendixA} and \ref{sec:AppendixB}. 
	
	\subsection{Bounding the Covariance Bottleneck via Restricted Predictor Overlap}
	\label{subsec:cov_bottleneck}
	
	The variance-reduction mechanism of an ensemble is fundamentally governed by the covariance between its constituent models \citep{ueda1996generalization}. Given a centered design matrix $\mathbf{X} \in \mathbb{R}^{n \times p}$ and $G$ linear models with coefficient vectors $\boldsymbol{\beta}^1, \dots, \boldsymbol{\beta}^G \in \mathbb{R}^p$, let $\mathbf{f}_g = \mathbf{X}\boldsymbol{\beta}^g$ denote the prediction vector for model $g$. The empirical variance of the ensemble prediction, $\bar{\mathbf{f}} = \frac{1}{G}\sum_{g=1}^G \mathbf{f}_g$, decomposes into the average individual model variance $\bar{V}_n$ and the average pairwise prediction covariance $\bar{C}_n$:
	\begin{equation}
		\mathrm{Var}_n(\bar{\mathbf{f}}) = \frac{1}{G}\bar{V}_n + \frac{G-1}{G}\bar{C}_n,
	\end{equation}
	where $\bar{V}_n = \frac{1}{G}\sum_{g=1}^G \mathrm{Var}_n(\mathbf{f}_g)$ and $\bar{C}_n = \frac{1}{G(G-1)}\sum_{g \neq h} \mathrm{Cov}_n(\mathbf{f}_g, \mathbf{f}_h)$. As the ensemble size $G$ grows, the first term vanishes, meaning the total prediction variance is bottlenecked by the average covariance $\bar{C}_n$. 
	
	To demonstrate how the MSS objective controls this bottleneck, we can establish a deterministic upper bound based on the variable overlap across the ensemble. Let $\mathbf{X}$ be standardized with unit diagonals and a maximum absolute off-diagonal correlation bounded by $r \in [0,1)$. Consider an ensemble where each of the $G$ constituent models is subject to the $\ell_1$ sparsity constraint $\|\boldsymbol{\beta}^g\|_1 \le s$. If we explicitly constrain the coefficient overlap between any pair of models by a continuous budget $\omega \ge 0$, such that $\sum_{j=1}^p |\beta_j^g \beta_j^h| \le \omega$ for all $g \neq h$, the empirical prediction variance of the ensemble is bounded from above:
	\begin{equation}
		\mathrm{Var}_n(\bar{\mathbf{f}}) \le \frac{1}{G}\bar{V}_n + \frac{G-1}{G} \Big[ \omega(1 - r) + r s^2 \Big].
	\end{equation}
	
	This bound isolates the effect of the splitting objective. Enforcing strictly disjoint feature subsets (achieved in exact MSS by setting $u=1$, which guarantees $\omega=0$) removes the $\mathcal{O}(1)$ covariance overlap, leaving only the unavoidable off-diagonal empirical correlation $r$. By penalizing $\omega$, the splitting framework mathematically forces $\bar{C}_n$ to shrink, ensuring that the ensemble variance continues to decrease as $G$ scales. We empirically validate this bound in Section \ref{sec:exact_empirical} by explicitly decomposing the bias, variance, and covariance of the MSS ensemble against the optimal single-model baseline.
	
	\subsection{Hedging Against Spurious Finite-Sample Correlations}
	\label{subsec:spurious_hedge}
	
	Beyond variance reduction, single sparse models are vulnerable to spurious empirical correlations. Finite-sample noise often causes uninformative variables to exhibit high sample correlations with true active predictors. We illustrate this vulnerability, and how splitting addresses it, using a stylized scenario.
	
	Assume a true linear model $Y = \beta^* X_1 + \varepsilon$, where $X_1 \sim \mathcal{N}(0,1)$ and $\varepsilon \sim \mathcal{N}(0, \sigma^2)$. Let $X_2 \sim \mathcal{N}(0,1)$ be an uninformative noise variable, completely independent of $X_1$ and $Y$ in the population ($\rho = 0$). Suppose that in a training sample of size $n$, $X_2$ exhibits a high spurious empirical correlation with $X_1$, such that $\mathrm{Corr}_n(\mathbf{x}_1, \mathbf{x}_2) = r \to 1$. 
	
	As $r \to 1$, the finite-sample probability of incorrectly selecting the spurious proxy $X_2$ via marginal screening approaches $0.5$. Consequently, on an independent, identically distributed test set, the Expected Prediction Error (EPE) of a single selected sparse model (such as Best Subset Selection) is inflated by this selection error, yielding a lower bound of:
	\begin{equation}
		\mathrm{EPE}_{\mathrm{BSS}} \ge \sigma^2 + (\beta^*)^2.
	\end{equation}
	
	Conversely, consider a disjoint Exact MSS ensemble ($G=2$) restricted to a maximum model size of $t=1$. Forced to build distinct models, the algorithm must select both the true signal and the spurious proxy. Letting $\hat{\beta}_1$ and $\hat{\beta}_2$ denote their respective ordinary least squares estimates, the ensemble prediction $\bar{f}(X) = \frac{1}{2}(\hat{\beta}_1 X_1 + \hat{\beta}_2 X_2)$ acts as a mathematical hedge. Averaging these models yields an expected prediction error bounded by:
	\begin{equation}
		\mathrm{EPE}_{\mathrm{MSS}} \approx \sigma^2 + \frac{1}{2}(\beta^*)^2.
	\end{equation}
	
	While simplified, this scenario isolates the mechanism by which single models fail when finite-sample noise generates spurious correlations, suffering generalization error proportional to the lost signal. By partitioning highly correlated variables across disjoint models, the split ensemble halves this worst-case excess risk. 
	
	This dynamic is particularly relevant in high-dimensional settings ($p \gg n$). With a massive candidate feature space, the probability of generating highly correlated spurious proxies naturally approaches 1 \citep{fan2016guarding}. Because a single sparse model is bound to encounter these proxies more frequently, its risk of generalization failure compounds. By actively distributing variables across $G$ sub-models, the ensemble provides multiple opportunities to capture the true causal signals alongside the proxies, continuously hedging against this recurring high-dimensional risk. We simulate this exact phenomenon in Section \ref{sec:exact_empirical} by explicitly controlling the empirical training correlation ($r$) via targeted data generation.
	
	% ---------------------------------------------------------
	\section{Empirical Validation of the Exact Oracle}
	\label{sec:exact_empirical}
	
	To demonstrate the statistical principles of Section \ref{sec:theory} free from approximation error, we evaluate the Exact MSS oracle \eqref{eq:exact_mss}. To permit an exhaustive search of all strictly disjoint partitions ($u=1$), we use a controlled simulation with $p=8$ predictors, artificially inducing spurious correlations to mimic high-dimensional traps. We benchmark exclusively against Exact BSS ($G=1$), which isolates pure feature selection without the shrinkage bias inherent to continuous relaxations like the Lasso or Elastic Net \citep{hastie2020best, hazimeh2020fast}.
	
	\subsection{Simulation Design}
	
	Responses are generated from the sparse linear model $\mathbf{y} = \mathbf{X}\boldsymbol{\beta}^* + \boldsymbol{\varepsilon}$,  where $\boldsymbol{\varepsilon} \sim \mathcal{N}(\mathbf{0}, \sigma^2 \mathbf{I}_n)$  is scaled to yield a Signal-to-Noise Ratio (SNR) of 1. The true coefficients $\boldsymbol{\beta}^* = (1, 1, 2, 2, 3, 3, 0, 0)^\top$ deliberately place active signals adjacent to uninformative noise. Letting $\boldsymbol{\Sigma}(\rho)$ denote an AR(1) covariance matrix with $\Sigma_{j,k} = \rho^{|j-k|}$, the experimental design imposes a deliberate asymmetry between training and test conditions:
	\begin{align*}
		\text{Training:} \quad 
		& \hat{\boldsymbol{\Sigma}}_{\text{train}} = 
		\boldsymbol{\Sigma}(r) \ \text{exactly}, 
		\quad r \in [0.1,\, 0.9], \quad n = 30, \\
		\text{Test:} \quad 
		& \mathbf{X}_{\text{test}} \sim 
		\mathcal{N}\!\left(\mathbf{0},\, 
		\boldsymbol{\Sigma}(\rho)\right), 
		\quad \rho = 0.1, \quad n_{\text{test}} = 5{,}000.
	\end{align*}
	Training sets are generated via the \texttt{simTargetCov} package \citep{R-simTargetCov} such that the sample covariance matrix  exactly equals $\boldsymbol{\Sigma}(r)$ (algorithmic details are 
	deferred to the Supplementary Material). When $r > \rho$, uninformative variables become highly correlated proxies for the true signals during training, a trap that severely magnifies as 
	$r \gg \rho$ \citep{fan2016guarding}. Because these spurious correlations are absent from the test set, any model misled by the training structure incurs severe generalization error. Exact MSS ($G=2$) and Exact BSS are computed via the \texttt{splitSelect} package \citep{R-splitSelect}.
	
	\subsection{Model Evaluation}
	
	Standard finite-sample cross-validation introduces severe selection variance at $n=30$, which can obscure the models' true predictive limits. To isolate the theoretical capacity of each objective function, we bypass cross-validation and exhaustively evaluate the expected out-of-sample performance for every valid combinatorial configuration. Let $S \in \mathcal{S}$ denote a candidate feature subset (for BSS) or disjoint partition (for MSS). For each configuration $S$, we compute the optimal coefficient vector $\hat{\boldsymbol{\beta}}_S^{(m)}$ on the $m$-th simulated training dataset under its respective objective.
	
	Each configuration $S$ is then evaluated on the independent test set $(\mathbf{X}_{\text{test}}, \mathbf{y}_{\text{test}})$, and  the Expected Prediction Error (EPE) is estimated by averaging  out-of-sample mean squared errors over $M = 500$ independent  training replications, scaled by the noise variance $\sigma^2$:
	\begin{equation}
		\label{eq:epe}
		\mathrm{EPE}_n(S) = \frac{1}{\sigma^2} \left\{ \frac{1}{M} \sum_{m=1}^M \left[ \frac{1}{n_{\text{test}}} \|\mathbf{y}_{\text{test}} - \mathbf{X}_{\text{test}} \hat{\boldsymbol{\beta}}_S^{(m)}\|_2^2 \right] \right\}.
	\end{equation}
	Scaling by $\sigma^2$ sets the theoretical lower bound (the Bayes 
	risk) to exactly 1. Computing $\mathrm{EPE}_n(S)$ for all 255 BSS 
	subsets and 3,025 MSS splits maps the complete risk landscape for 
	each framework, and the global minimum $\min_{S \in \mathcal{S}} 
	\mathrm{EPE}_n(S)$ yields the lowest attainable prediction error.

	% ---------------------------------------------------------
	\subsection{Expected Prediction Error}
	
\begin{figure}[h]
		\centering
		\includegraphics[width=0.75\textwidth]{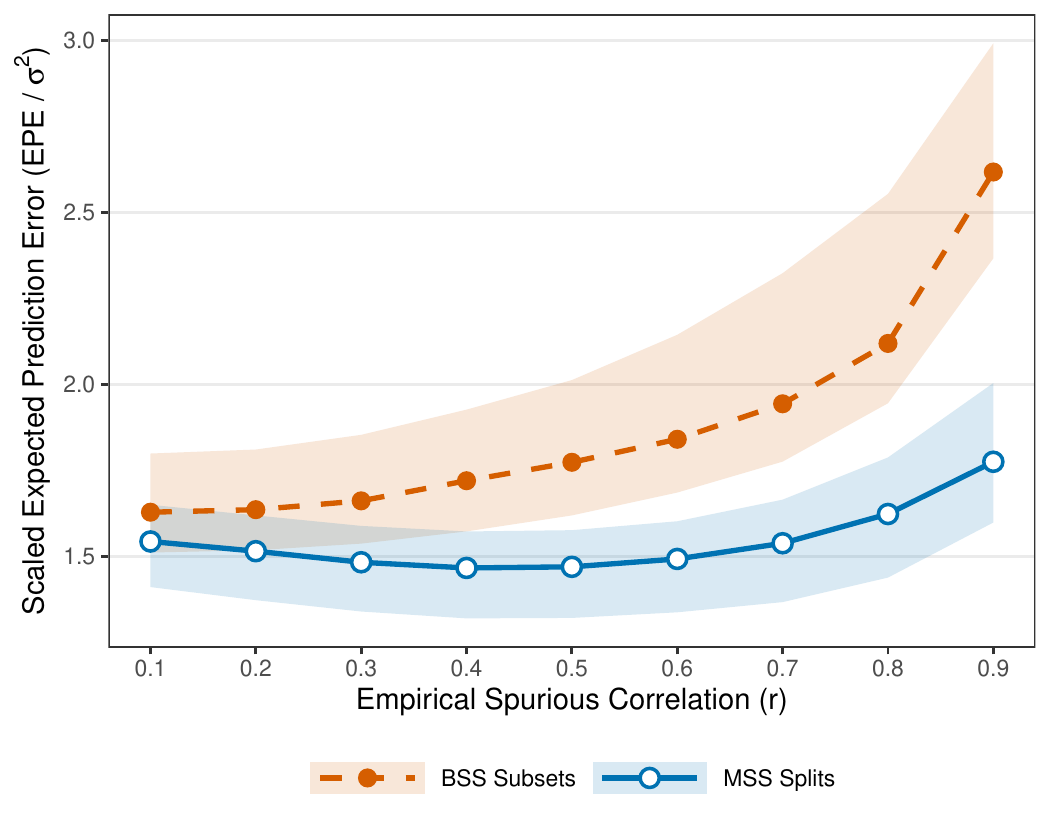}
		\caption{The scaled Expected Prediction Error (EPE) across varying levels of empirical spurious correlation $r$. Lines track the median EPE across all valid configurations, while shaded bands represent the interquartile range (25th to 75th percentiles).}
		\label{fig:mspe_r}
	\end{figure}

	Validating the mathematical hedge outlined in 
	Section~\ref{subsec:spurious_hedge}, Figure~\ref{fig:mspe_r} 
	tracks the median Expected Prediction Error (EPE) of all valid 
	subsets and splits across the training correlation grid, with 
	shaded bands representing the interquartile range (IQR). As $r$ 
	increases, single BSS models frequently discard valid signals, 
	relying instead on highly correlated proxies to minimize the 
	empirical training loss. When evaluated on the independent test 
	set where this artificial correlation is absent, this 
	misspecification leads to severe generalization error, with both 
	the median EPE and IQR of the BSS subsets inflating rapidly with 
	$r$.
	
	Conversely, the MSS objective explicitly restricts predictor overlap, forcing the ensemble to partition correlated features across different sub-models. By distributing the variables in this manner, the ensemble retains a broader proportion of the true active predictors. As evidenced by the relative stability of the median EPE and its tight IQR band in Figure \ref{fig:mspe_r}, the ensemble framework maintains stable predictive performance despite the high empirical correlation.
	
\begin{figure}[h]
		\centering
		\includegraphics[width=0.75\textwidth]{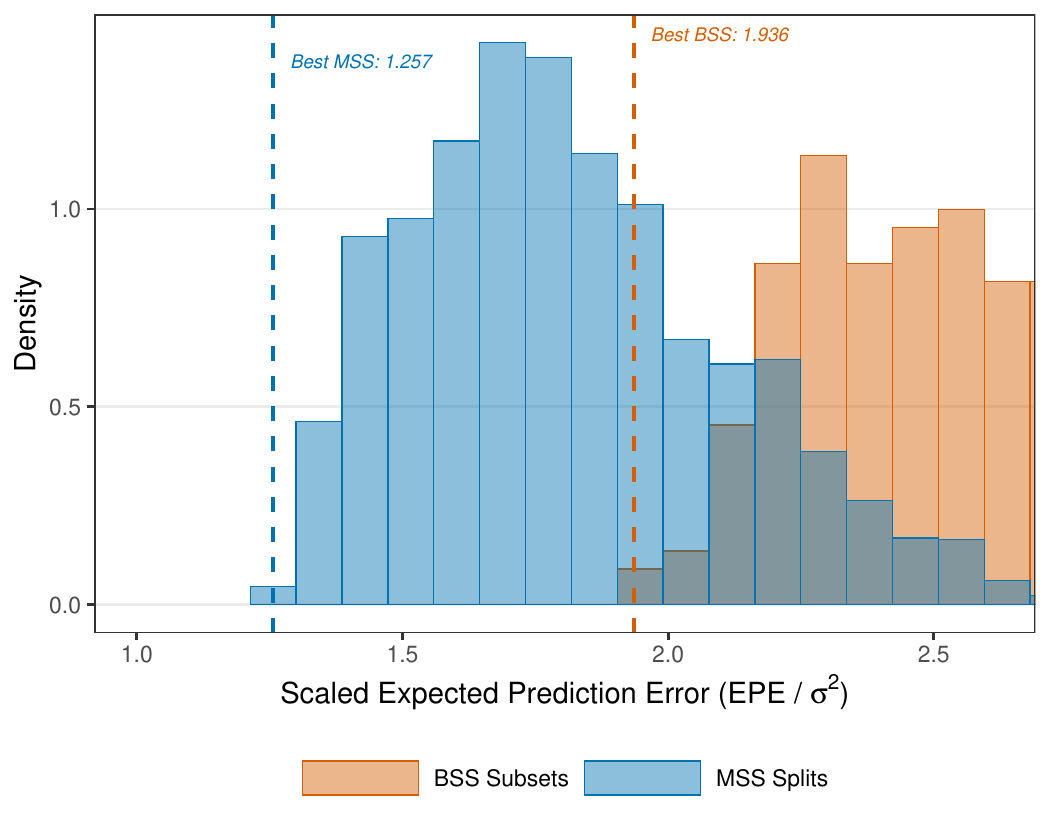}
		\caption{The relative density of expected prediction errors for all 3,025 valid MSS splits compared to all 255 valid BSS subsets at $r=0.90$. Vertical dashed lines highlight the absolute minimum (Oracle) EPE attainable by each framework.}
		\label{fig:risk_landscape}
	\end{figure}

	To demonstrate that this advantage extends across the entire model space, Figure \ref{fig:risk_landscape} plots the complete risk landscape for all valid configurations at the extreme correlation $r=0.90$. Vertical dashed lines denote the optimal configurations, showing that the best MSS ensemble ($\mathrm{EPE} = 1.257$) yields a substantially lower prediction risk than the best BSS subset ($\mathrm{EPE} = 1.936$). Notably, the bulk of the MSS risk distribution is shifted heavily toward lower error rates; over 68\% of \textit{all valid} MSS partitions naturally achieve a lower EPE than the absolute optimal BSS model. This indicates that the splitting objective systematically improves the risk landscape, making the ensemble approach highly resilient to the specific subset of variables selected.
	
	% ---------------------------------------------------------
	\subsection{Bias-Variance-Covariance Decomposition}
	
	To explicitly examine the statistical mechanics driving this reduction in prediction risk (Section \ref{subsec:cov_bottleneck}), we decompose the expected prediction error of the optimal configurations at $r = 0.90$ into squared bias, individual model variance, and pairwise prediction covariance.
	
	As detailed in Table \ref{tab:bias_var}, the constituent models within the optimal MSS ensemble exhibit relatively high individual variances (0.321), a natural consequence of their restricted, sparse feature sets. However, enforcing strictly disjoint partitions heavily suppresses their pairwise prediction covariance (0.047). This decorrelation restricts the overall ensemble variance to 0.184, a substantial reduction compared to the 0.729 variance of the optimal BSS model. Furthermore, because the objective distributes the candidate features across multiple equations, MSS simultaneously achieves a strictly lower squared bias (0.056) than BSS (0.206).
	
	\begin{table}[htbp]
		\centering
		\caption{Bias-Variance-Covariance decomposition of the lowest attainable Expected Prediction Error at empirical correlation $r = 0.90$. All components are scaled by the irreducible noise variance $\sigma^2$.}
		\vspace{0.2cm}
		\begin{tabular}{l ccccc}
			\toprule
			\textbf{Method} & \textbf{Bias}$^2$ & \textbf{Indiv. Var} & \textbf{Pairwise Cov} & \textbf{Total Var} & \textbf{Total EPE} \\
			\midrule
			Best BSS & 0.206 & --- & --- & 0.729 & 1.936 \\
			Best MSS & 0.056 & 0.321 & 0.047 & 0.184 & 1.257 \\
			\bottomrule
		\end{tabular}
		\label{tab:bias_var}
	\end{table}
	
	To illustrate how this simultaneous reduction in bias and variance is achieved, Table \ref{tab:allocation} maps the variable assignments of the respective optimal configurations. The true active signals in this simulation are $\{X_1, \dots, X_6\}$. Constrained by the high empirical collinearity, the optimal BSS model omits valid active signals ($X_2$ and $X_4$) to prevent severe variance inflation, relying instead on their highly correlated neighbors. 
	
	\begin{table}[htbp]
		\centering
		\caption{Feature allocation of the optimal BSS subset and MSS split at empirical correlation $r = 0.90$. Checkmarks denote the inclusion of a variable in the respective model. Variables $\mathbf{X_1}$ through $\mathbf{X_6}$ represent the true active signals, while $\mathbf{X_7}$ and $\mathbf{X_8}$ are uninformative noise variables.}
		\vspace{0.2cm}
		\renewcommand{\arraystretch}{1.5}
		\resizebox{0.9\textwidth}{!}{
			\begin{tabular}{l cccccc c cc}
				\toprule
				& \multicolumn{6}{c}{\textbf{True Active Signals}} && \multicolumn{2}{c}{\textbf{Noise Variables}} \\
				\cmidrule(lr){2-7} \cmidrule(lr){9-10}
				\textbf{Framework} & $\mathbf{X_1}$ & $\mathbf{X_2}$ & $\mathbf{X_3}$ & $\mathbf{X_4}$ & $\mathbf{X_5}$ & $\mathbf{X_6}$ && $\mathbf{X_7}$ & $\mathbf{X_8}$ \\
				\midrule
				\textbf{Best BSS Subset} & \cellcolor{orange!30} \checkmark & --- & \cellcolor{orange!30} \checkmark & --- & \cellcolor{orange!30} \checkmark & \cellcolor{orange!30} \checkmark && --- & --- \\
				\midrule
				\textbf{MSS Model 1} & --- & --- & \cellcolor{blue!20} \checkmark & --- & --- & \cellcolor{blue!20} \checkmark && --- & --- \\
				\textbf{MSS Model 2} & \cellcolor{cyan!20} \checkmark & --- & --- & \cellcolor{cyan!20} \checkmark & \cellcolor{cyan!20} \checkmark & --- && --- & --- \\
				\bottomrule
			\end{tabular}
		}
		\label{tab:allocation}
	\end{table}
	
	Given the extreme finite-sample limits ($r=0.90, n=30$), even the ensemble must omit $X_2$ to maintain variance control. However, the MSS framework relaxes the strict capacity constraints of a single sparse model. By partitioning correlated variables into separate equations, the ensemble is able to incorporate $X_4$. Because the true effect of $X_4$ ($\beta^*_4=2$) is double that of the omitted $X_2$ ($\beta^*_2=1$), isolating these features across two models recovers critical signal space without the variance penalty of forcing them into a single equation. This targeted recovery directly explains the sharp reduction in squared bias from 0.206 to 0.056.

	% ---------------------------------------------------------
	\section{Scaling Up: A Review of Tractable Split Modeling}
	\label{sec:review}
	
	While Exact MSS provides a theoretically optimal mechanism for variance reduction, its combinatorial nature renders it computationally intractable for high-dimensional data. To scale the objective-driven ensemble paradigm to modern applications, recent literature has developed a suite of computationally tractable algorithms. These approaches rely on continuous relaxations, projected gradient methods, and specialized loss functions to approximate the idealized splitting objective efficiently.
	
	The first major advancement in scaling this framework was the development of multi-convex relaxations, most notably Split Regularized Regression, or SplitReg \citep{christidis2020split}. To bypass the NP-hard $\ell_0$ constraints, SplitReg replaces the exact sparsity requirement with an Elastic Net penalty applied individually to each model. To enforce predictor diversity, the strict disjoint constraint ($u=0$) is relaxed into a continuous $\ell_1$-norm diversity penalty that penalizes the absolute product of coefficients across models. Because this relaxed objective function is multi-convex, it can be minimized efficiently using block coordinate descent algorithms, allowing the splitting framework to be applied directly to high-dimensional datasets.
	
	More recently, advancements in optimization have allowed the framework to return to exact $\ell_0$-constrained formulations without relying on exhaustive combinatorial searches. Fast Multi-Model Subset Selection \citep{christidis2025multi} approximates the Exact MSS objective using projected gradient algorithms. By employing Iterative Hard Thresholding, this approach simultaneously updates the ensemble coefficients and projects them onto the restricted $\ell_0$ sparsity and diversity spaces. This provides a highly scalable alternative that explicitly preserves the hard variable-dropping properties of the exact oracle.
	
	Because the generalized MSS objective in \eqref{eq:exact_mss} is defined over an arbitrary empirical loss function $\mathcal{L}(\cdot, \cdot)$, the framework is highly modular and naturally extends beyond continuous regression. By substituting the squared error loss with a logistic loss, the framework was adapted to Multi-Model Logistic Regression \citep{christidis2025data}, yielding highly interpretable classification ensembles. This methodology was further generalized by \citet{yang2026diverse}, who incorporated cost-sensitive loss functions to handle highly imbalanced datasets, demonstrating that objective-driven splitting maintains its predictive superiority in complex classification settings.
	
	Finally, the splitting framework has been extended to robust statistics to handle contaminated data. To protect against \textit{casewise} contamination, where entire observations are corrupted, Robust Multi-Model Subset Selection \citep{christidis2026robust} adapts the MSS objective for robust loss functions via fast projected gradient updates. Conversely, high-dimensional datasets increasingly suffer from \textit{cellwise} contamination, where individual data entries are corrupted independently. Because joint optimization is computationally prohibitive under cellwise noise, the splitting principles have been adapted into a sequential approximation \citep{christidis2026fast}. Utilizing cellwise-robust estimators, this algorithm iteratively partitions features into disjoint subsets, successfully recovering the variance-reduction benefits of MSS. Together, these robust generalizations allow objective-driven ensembles to be reliably deployed across a wide variety of challenging domains.
	
	% ---------------------------------------------------------
	\section{High-Dimensional Benchmarking}
	\label{sec:high_dim}
	
	While the exact combinatorial oracle in Section \ref{sec:exact_empirical} isolates the statistical mechanics of the splitting objective, it is computationally restricted to trivial dimensions. In this section, we evaluate whether these theoretical advantages hold in a modern, high-dimensional setting ($p \gg n$) using the tractable computational approximations reviewed in Section \ref{sec:review}.
	
	\subsection{Simulation Design}
	
	We construct a high-dimensional regression setting with $n=50$ training observations and $p=500$ candidate predictors, yielding a dimensionality ratio of $p/n = 10$. To accurately reflect the statistical challenges inherent to modern applications, such as genomics (where genes act in co-expressed pathways) or quantitative finance (where asset returns are heavily sector-correlated), the design matrix is generated with a highly correlated block-diagonal covariance structure. 
	
	Specifically, the $500$ predictors are divided into $20$ independent blocks of $25$ variables each. Within each block, variables exhibit an autoregressive AR(1) correlation structure with a severe correlation coefficient of $\rho = 0.90$. The true signal is dense but highly localized: exactly $100$ predictors are active, distributed entirely within $4$ of the $20$ blocks. The non-zero coefficients are generated with alternating signs and varying magnitudes. The irreducible error variance ($\sigma^2$) is scaled to yield a challenging SNR of 1, which corresponds to a Proportion of Variance Explained (PVE) of exactly 50\%. This localized, highly collinear, and noisy structure creates an exceptionally difficult variable selection task, as models must disentangle true signals from highly correlated intra-block noise proxies without overfitting the limited sample size.
	
	\subsection{Benchmark Methods}
	
	We evaluate how effectively different methodologies navigate the trade-off between prediction risk and support recovery under severe finite-sample collinearity. To establish comprehensive baselines, we select algorithms from opposite ends of the modeling spectrum, contrasting interpretable sparse penalties with purely predictive algorithmic ensembles. Specifically, we benchmark three distinct philosophies:
	
	\begin{itemize}
		\item \textbf{Single-model sparse regularizers:} The Lasso and the Elastic Net, computed via the \texttt{glmnet} package \citep{R-glmnet}, alongside Fast BSS, computed via the \texttt{L0Learn} package \citep{R-L0Learn}.
		
		\item \textbf{Objective-driven ensembles:} The methods reviewed in Section \ref{sec:review}, specifically SplitReg, computed via the \texttt{SplitGLM} package \citep{R-SplitGLM}, and Fast MSS, computed via the \texttt{PSGD} package \citep{R-PSGD}. Both are configured to construct $G=5$ sparse sub-models.
		
		\item \textbf{Algorithmic black-box ensembles:} Random Forests \citep{breiman2001random}, computed via the \texttt{ranger} package \citep{R-ranger}; Random Generalized Linear Models (RGLM) \citep{song2013random}, computed via the \texttt{randomGLM} package \citep{R-randomGLM}; and Gradient Boosting, computed via the \texttt{xgboost} package \citep{chen2016xgboost}.
	\end{itemize}
	
	To explicitly assess the impact of ensemble size on the algorithmic black-box methods, Random Forests and RGLM are evaluated both at their default sizes ($G=500$ and $G=100$, respectively) and at a restricted size of $G=5$ to match the capacity of the objective-driven ensembles. All relevant penalty parameters for the sparse and objective-driven methods are selected via 5-fold cross-validation on the training set.
	
	\subsection{Performance Measures}
	
	All models are evaluated on independently generated test sets of $n_{\text{test}} = 5{,}000$ observations. Predictive accuracy is assessed via the scaled Mean Squared Prediction Error (MSPE):
	\begin{equation}
		\mathrm{MSPE} = \frac{1}{\sigma^2} \left\{ \frac{1}{n_{\text{test}}} \|\mathbf{y}_{\text{test}} - \hat{\mathbf{y}}\|_2^2 \right\},
	\end{equation}
	which normalizes the theoretical lower bound (the Bayes risk) to exactly 1.
	
	To evaluate interpretability and variable selection, let $\mathcal{T}$ denote the set of true active predictors and $\widehat{\mathcal{A}}$ denote the set of predictors utilized by the fitted model. For ensembles, $\widehat{\mathcal{A}}$ is defined as the union of all features used across constituent sub-models. We evaluate selection accuracy using Precision (PR) and Recall (RC):
	\begin{equation}
		\mathrm{PR} = \frac{|\widehat{\mathcal{A}} \cap \mathcal{T}|}{|\widehat{\mathcal{A}}|}, \quad \quad \mathrm{RC} = \frac{|\widehat{\mathcal{A}} \cap \mathcal{T}|}{|\mathcal{T}|}.
	\end{equation}
	Precision and recall, both bounded in $[0,1]$, quantify the proportion of selected features that are true signals and the fraction of true signals successfully discovered, respectively. A precision of $1$ implies zero false positives, while a recall of $1$ implies zero false negatives. To summarize this trade-off, we compute the F1 Score, defined as their harmonic mean: $\mathrm{F1} = 2 \cdot (\mathrm{PR} \cdot \mathrm{RC}) / (\mathrm{PR} + \mathrm{RC})$. Also bounded in $[0, 1]$, an F1 Score of $1$ denotes perfect support recovery (selecting exactly the true active predictors and zero noise variables), whereas $0$ denotes a complete failure to identify any active features. Thus, a high F1 Score indicates an interpretable model that captures the underlying signal while strictly filtering out spurious noise. All reported metrics are averaged over 50 independently generated datasets.
	
	\subsection{Results}
	
\begin{figure}[h]
		\centering
		\includegraphics[width=0.75\textwidth]{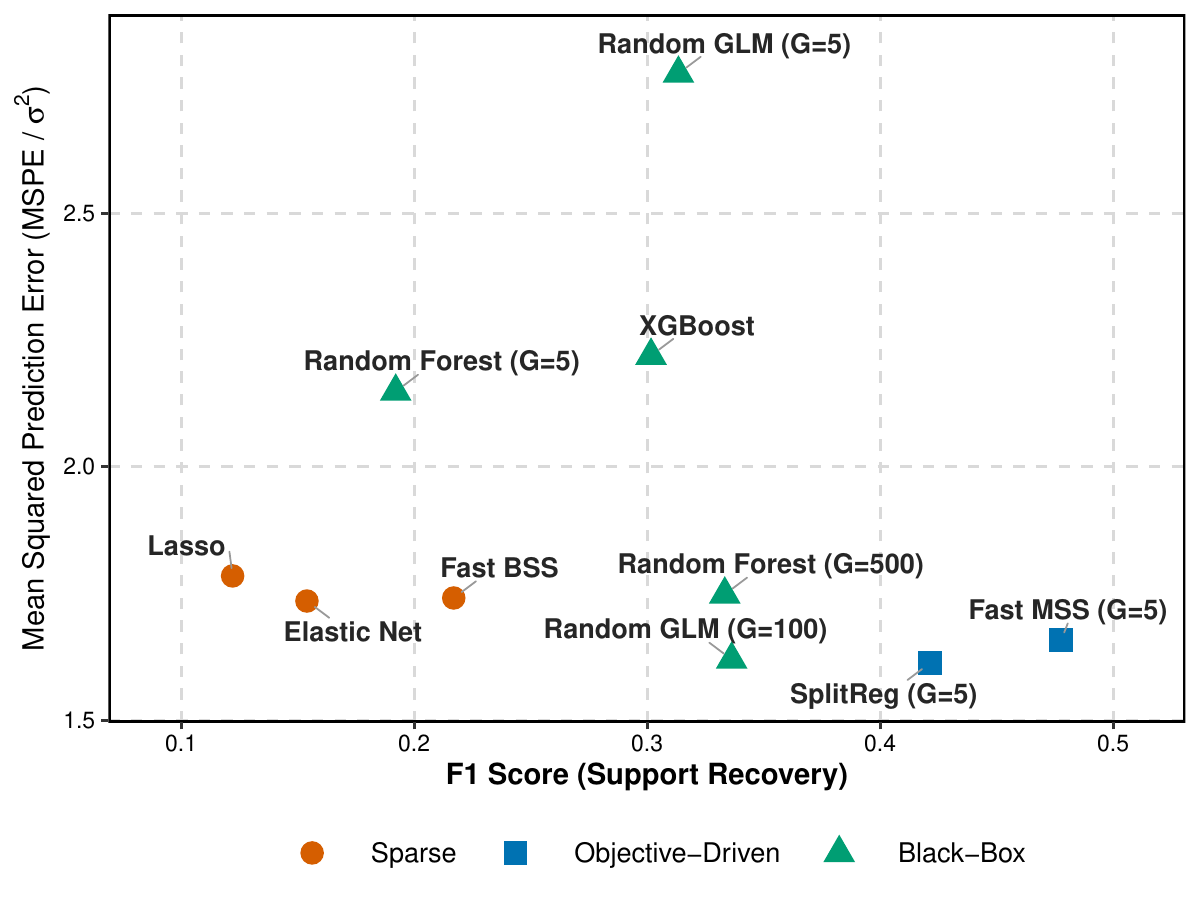}
		\caption{The Pareto frontier of predictive risk versus support recovery (F1 Score) in the high-dimensional simulation. The optimal performance region is the bottom-right quadrant.}
		\label{fig:pareto}
	\end{figure}

	Table \ref{tab:high_dim_results} summarizes the performance metrics across the evaluated methods, while Figure \ref{fig:pareto} visualizes the inherent trade-off between predictive risk and support recovery. 
	
	The results clearly illustrate the fundamental divide between single sparse models and massive algorithmic ensembles. Single sparse regularizers (Lasso, Elastic Net, Fast BSS) achieve strong precision (0.55 to 0.60), successfully isolating true signals. However, severe $p \gg n$ collinearity forces high false negative rates, keeping recall scores below 0.18. Missing these broader signal pathways elevates prediction risk ($\mathrm{EPE} \approx 1.75$). Conversely, massive black-box ensembles like Random GLM ($G=100$) achieve excellent prediction accuracy ($\mathrm{EPE} = 1.620$) but completely abandon interpretability. Because exactly 100 of the 500 candidate predictors are active, selecting nearly all available features inherently drives precision down to the baseline rate of 0.20. Attempting to force interpretability onto these algorithmic methods by restricting their size ($G=5$) leads to severe prediction failure ($\mathrm{EPE} > 2.1$). Inherently reliant on massive randomization to stabilize variance, they cannot function effectively as small, interpretable ensembles.

	\begin{table}[!htbp]
		\centering
		\caption{High-dimensional simulation results averaged over 50 replications ($n=50, p=500$). }
		\vspace{0.2cm}
		\begin{tabular}{lccccc}
			\toprule
			\textbf{Method} && &\textbf{MSPE} & \textbf{RC} & \textbf{PR} \\
			\midrule
			\addlinespace[0.6em]
			\multicolumn{4}{l}{\textbf{Single Sparse Models}} \\
			\addlinespace[0.3em]
			\hspace{1em}Elastic Net &&& 1.735 & 0.096 & 0.559 \\
			\hspace{1em}Fast BSS &&& 1.741 & 0.176 & 0.601 \\
			\hspace{1em}Lasso &&& 1.785 & 0.075 & 0.555 \\
			\addlinespace[0.6em]
			\multicolumn{4}{l}{\textbf{Objective-Driven Ensembles}} \\
			\addlinespace[0.3em]
			\hspace{1em}SplitReg ($G=5$) &&& 1.613 & 0.679 & 0.351 \\
			\hspace{1em}Fast MSS ($G=5$) &&& 1.658 & 0.432 & 0.593 \\
			\addlinespace[0.6em]
			\multicolumn{4}{l}{\textbf{Black-Box Ensembles}} \\
			\addlinespace[0.3em]
			\hspace{1em}Random GLM ($G=5$) &&& 2.775 & 0.358 & 0.279 \\
			\hspace{1em}Random GLM ($G=100$) &&& 1.620 & 0.996 & 0.202 \\
			\hspace{1em}Random Forest ($G=5$) &&& 2.148 & 0.163 & 0.235 \\
			\hspace{1em}Random Forest ($G=500$) &&& 1.748 & 1.000 & 0.200 \\
			\hspace{1em}XGBoost &&& 2.219 & 0.434 & 0.231 \\
			\bottomrule
		\end{tabular}
		\label{tab:high_dim_results}
	\end{table}

	Objective-driven ensembles successfully bridge this divide, occupying the optimal bottom-right quadrant in Figure \ref{fig:pareto}. SplitReg achieves the lowest prediction risk ($\mathrm{MSPE} = 1.613$), while Fast MSS maintains strict sparsity to yield the highest support recovery ($\mathrm{F1} = 0.478$). The mechanics driving this performance are evident in the individual sub-models. An isolated SplitReg sub-model yields a prediction error ($\mathrm{MSPE} = 1.641$) comparable to the standalone Elastic Net ($\mathrm{MSPE} = 1.735$). Similarly, a single Fast MSS sub-model matches the precision of Fast BSS (0.599 versus 0.601) but captures few signals, averaging an individual $\mathrm{MSPE}$ of 2.194. By penalizing overlap, the joint objective forces these sparse sub-models to explore distinct collinear blocks. Aggregating these decorrelated models rescues lost signal space and neutralizes finite-sample variance, driving the Fast MSS ensemble $\mathrm{MSPE}$ down to 1.658. This illustrates how joint optimization engineers the variance reduction single models lack while retaining strict interpretability.

	% ---------------------------------------------------------
	\section{Discussion}
	\label{sec:discussion}
	
	The dichotomy between interpretable sparse models and black-box ensembles is not an inherent mathematical necessity. By formulating ensemble generation as a joint optimization problem, the objective-driven splitting paradigm rigorously bridges this gap. As demonstrated through finite-sample mechanics and validated empirically, actively penalizing predictor overlap controls prediction covariance and mitigates the spurious sample correlations that often compromise single sparse models. Scaled via fast combinatorial algorithms and multi-convex relaxations, this framework extracts the variance reduction benefits of ensembles while preserving the strict global sparsity required for reliable interpretation.
	
	It is important to distinguish the goals of objective-driven ensembles from classical variable selection consistency. In highly correlated, high-dimensional environments, identifying the exact true causal mechanism is often statistically impossible without strong, untestable assumptions. Single sparse models often project a false sense of certainty by arbitrarily selecting one proxy over another, leaving predictions vulnerable to severe generalization error. By explicitly penalizing predictor overlap, the splitting framework prioritizes predictive stability over strict single-model selection. It transparently presents the user with a small set of alternative, mathematically valid explanatory pathways, functioning as a systematic safeguard against model non-identifiability.
	
	To facilitate widespread adoption, a comprehensive suite of scalable algorithms is publicly available on CRAN. This ecosystem includes \texttt{SplitReg} \citep{R-SplitReg} and \texttt{SplitGLM} \citep{R-SplitGLM} for multi-convex relaxations in linear and generalized linear models, alongside \texttt{PSGD} \citep{R-PSGD} for fast $\ell_0$-constrained Multi-Model Subset Selection. For contaminated environments, the \texttt{RMSS} \citep{R-RMSS} and \texttt{srlars} \citep{R-srlars} packages provide fast, robust ensemble implementations for casewise and cellwise outliers, respectively. However, these implementations do not represent an exhaustion of what can be achieved with this methodology. Rather, they serve as foundational building blocks, marking the beginning of a broader paradigm shift in how ensemble modeling can be approached mathematically rather than heuristically. 
	
	For example, this paradigm is not limited to generalized linear models. A natural progression is the development of objective-driven tree ensembles. Rather than relying on heuristic bootstrap aggregating or random feature subspacing to generate diversity, future work could jointly optimize a small ensemble of shallow decision trees subject to explicit constraints on predictor overlap. This extension, alongside future adaptations for survival analysis and dynamic prediction, would bring the rigorous mathematical benefits of the splitting framework to an even wider array of statistical learning tasks.
	
	\section*{Data and Code Availability}
	
	All source code, simulation scripts, and data generation procedures required to fully reproduce the results, tables, and figures presented in this manuscript are publicly available in the GitHub repository at \url{https://github.com/AnthonyChristidis/objective-driven-ensembles}.
	
	\section*{Conflicts of Interest}
	
	The authors declare no competing interests.
	
	\appendix
	
	\titleformat{\section}{\Large\bfseries}{Appendix \thesection:}{0.5em}{}

	\section{Derivation of the Covariance Bound (Section \ref{subsec:cov_bottleneck})} \label{sec:AppendixA}
	
	This section provides the detailed algebraic derivation for the deterministic upper bound on the empirical prediction variance presented in Section 3.1 of the main manuscript.
	
	\textbf{Setup:} Let $\mathbf{X} \in \mathbb{R}^{n \times p}$ be a standardized design matrix with empirical covariance $\hat{\mathbf{\Sigma}} = \frac{1}{n}\mathbf{X}^\top \mathbf{X}$, unit diagonals, and maximum absolute off-diagonal correlation bounded by $r \in [0,1)$. Consider an ensemble of $G$ linear models defined by coefficient vectors $\boldsymbol{\beta}^g = (\beta_1^g, \dots, \beta_p^g)^\top$, each subject to the $\ell_1$ sparsity constraint $\|\boldsymbol{\beta}^g\|_1 \le s$. The coefficient overlap between any pair of models is constrained by a continuous budget $\omega \ge 0$ such that $\sum_{j=1}^p |\beta_j^g \beta_j^h| \le \omega$ for all $g \neq h$. We aim to show that the empirical prediction variance of the ensemble $\bar{\mathbf{f}}$ is bounded from above by:
	\[
	\mathrm{Var}_n(\bar{\mathbf{f}}) \le \frac{1}{G}\bar{V}_n + \frac{G-1}{G} \Big[ \omega(1 - r) + r s^2 \Big].
	\]
	
	\textbf{Derivation:}
	Let $\mathbf{f}_g = \mathbf{X}\boldsymbol{\beta}^g$ and $\mathbf{f}_h = \mathbf{X}\boldsymbol{\beta}^h$ represent the predictions for any two distinct models $g \neq h$. The empirical covariance between these prediction vectors is given by the quadratic form:
	\[
	\mathrm{Cov}_n(\mathbf{f}_g, \mathbf{f}_h) = \frac{1}{n} (\mathbf{X}\boldsymbol{\beta}^g)^\top (\mathbf{X}\boldsymbol{\beta}^h) = (\boldsymbol{\beta}^g)^\top \hat{\mathbf{\Sigma}} \boldsymbol{\beta}^h.
	\]
	We decompose this sum into its diagonal and off-diagonal components:
	\[
	(\boldsymbol{\beta}^g)^\top \hat{\mathbf{\Sigma}} \boldsymbol{\beta}^h = \sum_{j=1}^p \hat{\mathbf{\Sigma}}_{jj} \beta_j^g \beta_j^h + \sum_{j \neq k} \hat{\mathbf{\Sigma}}_{jk} \beta_j^g \beta_k^h.
	\]
	Since the predictors are standardized ($\hat{\mathbf{\Sigma}}_{jj} = 1$), we apply the triangle inequality to bound the absolute covariance:
	\[
	\big|\mathrm{Cov}_n(\mathbf{f}_g, \mathbf{f}_h)\big| \le \sum_{j=1}^p \big|\beta_j^g \beta_j^h\big| + \sum_{j \neq k} |\hat{\mathbf{\Sigma}}_{jk}| \big|\beta_j^g\big| \big|\beta_k^h\big|.
	\]
	By the assumption that the maximum pairwise empirical correlation is bounded by $r$, we have:
	\[
	\big|\mathrm{Cov}_n(\mathbf{f}_g, \mathbf{f}_h)\big| \le \sum_{j=1}^p \big|\beta_j^g \beta_j^h\big| + r \sum_{j \neq k} \big|\beta_j^g\big| \big|\beta_k^h\big|.
	\]
	Next, we utilize the algebraic expansion for the product of the $\ell_1$ norms of the two vectors:
	\[
	\|\boldsymbol{\beta}^g\|_1 \|\boldsymbol{\beta}^h\|_1 = \left( \sum_{j=1}^p \big|\beta_j^g\big| \right) \left( \sum_{k=1}^p \big|\beta_k^h\big| \right) = \sum_{j=1}^p \big|\beta_j^g \beta_j^h\big| + \sum_{j \neq k} \big|\beta_j^g\big| \big|\beta_k^h\big|.
	\]
	Rearranging this identity to isolate the off-diagonal sum yields:
	\[
	\sum_{j \neq k} \big|\beta_j^g\big| \big|\beta_k^h\big| = \|\boldsymbol{\beta}^g\|_1 \|\boldsymbol{\beta}^h\|_1 - \sum_{j=1}^p \big|\beta_j^g \beta_j^h\big|.
	\]
	Substituting this back into the covariance bound gives:
	\begin{align*}
		\big|\mathrm{Cov}_n(\mathbf{f}_g, \mathbf{f}_h)\big| &\le \sum_{j=1}^p \big|\beta_j^g \beta_j^h\big| + r \left( \|\boldsymbol{\beta}^g\|_1 \|\boldsymbol{\beta}^h\|_1 - \sum_{j=1}^p \big|\beta_j^g \beta_j^h\big| \right) \\
		&= (1 - r) \sum_{j=1}^p \big|\beta_j^g \beta_j^h\big| + r \|\boldsymbol{\beta}^g\|_1 \|\boldsymbol{\beta}^h\|_1.
	\end{align*}
	Applying the coefficient overlap constraint $\sum |\beta_j^g \beta_j^h| \le \omega$ and the sparsity constraints $\|\boldsymbol{\beta}^g\|_1 \|\boldsymbol{\beta}^h\|_1 \le s^2$, we obtain the strict upper bound on the pairwise covariance:
	\[
	\mathrm{Cov}_n(\mathbf{f}_g, \mathbf{f}_h) \le \big|\mathrm{Cov}_n(\mathbf{f}_g, \mathbf{f}_h)\big| \le \omega(1 - r) + r s^2.
	\]
	Finally, recalling from Equation (1) in the main manuscript that the empirical variance of the ensemble prediction $\bar{\mathbf{f}}$ decomposes into the average individual model variance $\bar{V}_n$ and the average pairwise covariance $\bar{C}_n$, substituting the upper bound for the covariance terms yields:
	\[
	\mathrm{Var}_n(\bar{\mathbf{f}}) = \frac{1}{G}\bar{V}_n + \frac{G-1}{G}\bar{C}_n \le \frac{1}{G}\bar{V}_n + \frac{G-1}{G} \Big[ \omega(1 - r) + r s^2 \Big].
	\]
	
	\section{Derivation of the Mathematical Hedge against Spurious Correlations (Section \ref{subsec:spurious_hedge})} \label{sec:AppendixB}
	
	This section provides the detailed mathematical calculations demonstrating how the Exact MSS framework provides a mathematical hedge against spurious sample correlations, as discussed in Section 3.2 of the main manuscript.
	
	\textbf{Setup:} Assume a true linear model $Y = \beta^* X_1 + \varepsilon$, where $X_1 \sim \mathcal{N}(0,1)$ and $\varepsilon \sim \mathcal{N}(0, \sigma^2)$. Let $X_2 \sim \mathcal{N}(0, 1)$ be an uninformative noise variable, completely independent of $X_1$ and $Y$ in the population ($\rho = 0$). Suppose that in a training sample of size $n$, $X_2$ exhibits a high spurious empirical correlation with $X_1$, such that $\mathrm{Corr}_n(\mathbf{x}_1, \mathbf{x}_2) = r \to 1$. We aim to show that as $r \to 1$, the finite-sample probability of incorrectly selecting the spurious proxy $X_2$ via marginal screening approaches $0.5$, bottlenecking the Expected Prediction Error (EPE) of a single sparse model at $\sigma^2 + (\beta^*)^2$. We then demonstrate how an Exact MSS ensemble ($G=2, t=1$) mathematically halves this excess risk.
	
	\vspace{1em}
	\noindent \textbf{Part 1: Probability of Selection Failure.} \\
	Consider a marginal screener that evaluates the empirical covariance of each predictor with the response vector $\mathbf{y} = \beta^* \mathbf{x}_1 + \boldsymbol{\varepsilon}$. Let $Z_1$ and $Z_2$ denote these empirical covariances. Assuming the predictors are standardized to unit empirical variance ($\mathrm{Var}_n(\mathbf{x}_1) = 1$) with empirical correlation $\mathrm{Cov}_n(\mathbf{x}_1, \mathbf{x}_2) = r$, we have:
	\begin{align*}
		Z_1 &= \mathrm{Cov}_n(\mathbf{x}_1, \mathbf{y}) = \beta^* \mathrm{Var}_n(\mathbf{x}_1) + \frac{1}{n}\mathbf{x}_1^\top \boldsymbol{\varepsilon} = \beta^* + \frac{1}{n}\mathbf{x}_1^\top \boldsymbol{\varepsilon}, \\
		Z_2 &= \mathrm{Cov}_n(\mathbf{x}_2, \mathbf{y}) = \beta^* \mathrm{Cov}_n(\mathbf{x}_1, \mathbf{x}_2) + \frac{1}{n}\mathbf{x}_2^\top \boldsymbol{\varepsilon} = \beta^* r + \frac{1}{n}\mathbf{x}_2^\top \boldsymbol{\varepsilon}.
	\end{align*}
	The algorithm incorrectly selects $X_2$ if $Z_2 > Z_1$. Define the difference $\Delta = Z_1 - Z_2$:
	\[
	\Delta = \beta^*(1 - r) + \frac{1}{n}(\mathbf{x}_1 - \mathbf{x}_2)^\top \boldsymbol{\varepsilon}.
	\]
	Conditional on the design matrix, $\boldsymbol{\varepsilon}$ is a Gaussian vector. Therefore, $\Delta$ is a normally distributed random variable. Its conditional mean is $\mathbb{E}[\Delta | \mathbf{X}] = \beta^*(1 - r)$. Its conditional variance relies on the empirical variance of the difference between the two predictors:
	\[
	\mathrm{Var}(\Delta | \mathbf{X}) = \frac{\sigma^2}{n^2} \|\mathbf{x}_1 - \mathbf{x}_2\|_2^2 = \frac{\sigma^2}{n} \mathrm{Var}_n(\mathbf{x}_1 - \mathbf{x}_2) = \frac{2\sigma^2}{n}(1 - r).
	\]
	The probability of incorrectly selecting $X_2$ is $P(\Delta < 0)$. Standardizing this normal variable yields:
	\[
	P(\Delta < 0) = \Phi\left( \frac{0 - \beta^*(1 - r)}{\sqrt{\frac{2\sigma^2}{n}(1 - r)}} \right) = \Phi\left( -\beta^* \sqrt{\frac{n(1 - r)}{2\sigma^2}} \right),
	\]
	where $\Phi(\cdot)$ is the standard normal cumulative distribution function. As the spurious correlation becomes near-perfect ($r \to 1$), the term $(1-r) \to 0$. Consequently, the argument to $\Phi$ approaches 0, and the failure probability $P(\Delta < 0) \to \Phi(0) = 0.5$.
	
	\vspace{1em}
	\noindent \textbf{Part 2: EPE under Independent Test Data.} \\
	We evaluate the models on an independent test set where $X_1, X_2 \overset{iid}{\sim} \mathcal{N}(0, 1)$, breaking the spurious correlation. Because $r \approx 1$ during training, the ordinary least squares coefficient for whichever variable is selected approaches $\beta^*$ (i.e., $\hat{\beta}_1 \approx \beta^*$ and $\hat{\beta}_2 \approx \beta^*$).
	
	For the single model, with probability $\frac{1}{2}$ it selects the true variable $X_1$, yielding the optimal Bayes risk: $\mathbb{E}[(Y - \beta^* X_1)^2] = \mathbb{E}[\varepsilon^2] = \sigma^2$. With probability $\frac{1}{2}$ it selects the spurious variable $X_2$, which is independent of the true response $Y = \beta^* X_1 + \varepsilon$. The prediction error in this failure case is:
	\[
	\mathbb{E}\left[(Y - \beta^* X_2)^2\right] = \mathbb{E}\left[(\beta^* X_1 + \varepsilon - \beta^* X_2)^2\right] = (\beta^*)^2 \mathbb{E}[X_1^2] + \mathbb{E}[\varepsilon^2] + (\beta^*)^2 \mathbb{E}[X_2^2] = 2(\beta^*)^2 + \sigma^2.
	\]
	Averaging over the two selection outcomes, the expected risk of the single model is:
	\[
	\mathrm{EPE}_{\mathrm{BSS}} \approx \frac{1}{2}(\sigma^2) + \frac{1}{2}\left(2(\beta^*)^2 + \sigma^2\right) = \sigma^2 + (\beta^*)^2.
	\]
	
	Now consider the disjoint Exact MSS ensemble ($G=2, t=1$). By forcing a partition of the two correlated variables, the ensemble prediction relies equally on both: $\bar{f}(X) = \frac{1}{2}(\hat{\beta}_1 X_1 + \hat{\beta}_2 X_2) \approx \frac{1}{2}\beta^* X_1 + \frac{1}{2}\beta^* X_2$. The EPE on the test set is:
	\[
	\mathrm{EPE}_{\mathrm{MSS}} = \mathbb{E}\left[ \left( \beta^* X_1 + \varepsilon - \frac{1}{2}\beta^* X_1 - \frac{1}{2}\beta^* X_2 \right)^2 \right] = \mathbb{E}\left[ \left( \frac{1}{2}\beta^* X_1 - \frac{1}{2}\beta^* X_2 + \varepsilon \right)^2 \right].
	\]
	Because $X_1, X_2$, and $\varepsilon$ are mutually independent in the test set, the cross-terms vanish:
	\begin{align*}
		\mathrm{EPE}_{\mathrm{MSS}} &= \frac{1}{4}(\beta^*)^2 \mathbb{E}[X_1^2] + \frac{1}{4}(\beta^*)^2 \mathbb{E}[X_2^2] + \mathbb{E}[\varepsilon^2] \\
		&= \frac{1}{4}(\beta^*)^2 + \frac{1}{4}(\beta^*)^2 + \sigma^2 = \sigma^2 + \frac{1}{2}(\beta^*)^2.
	\end{align*}
	Thus, by explicitly enforcing a disjoint partition of the correlated variables, the ensemble strictly halves the worst-case excess risk caused by the spurious sample correlation.
	
	\section{Generating Data with Targeted Empirical Covariances} \label{sec:appendixC}
	
	To explicitly study the effect of finite-sample spurious correlations in Section~\ref{sec:exact_empirical} of the main manuscript, we generate training data that exhibits a specific, targeted empirical covariance structure. We consider data that follow a Gaussian distribution, although the underlying methodology can be applied to other distributions.
	
	Suppose that the data come from a population with mean vector $\mathbf{0}_p \in \mathbb{R}^p$ and population covariance matrix $\Sigma_{\rho} \in \mathbb{R}^{p \times p}$. To mimic the spurious correlation phenomenon observed in high-dimensional settings, we use spectral decomposition to modify the design matrix $\mathbf{X}$ of the training data such that it achieves an exact target empirical covariance matrix $\Sigma_r$. This empirical covariance may differ significantly from the underlying population covariance $\Sigma_{\rho}$.
	
	The procedure to generate samples with a target empirical covariance, based on a specific rotation utilizing the spectral decomposition of the target covariance matrix, is detailed in Algorithm~\ref{algo:spurious}. This procedure is implemented computationally in the \texttt{simTargetCov}  package.
	
	\begin{algorithm}[h!] 
		\caption{Generate Data with Target Empirical Covariance Matrix}
		\begin{algorithmic}[1]  
			\Require Population covariance matrix $\Sigma_{\rho} \in \mathbb{R}^{p \times p}$ and target sample covariance matrix $\Sigma_r \in \mathbb{R}^{p \times p}$.
			
			\State Generate the rows of the initial design matrix $\widetilde{\mathbf{X}}$ from $\mathcal{N}(\mathbf{0}_p, \Sigma_{\rho})$.
			\State Let $\widetilde{\mathbf{Z}}$ be the standardized version of $\widetilde{\mathbf{X}}$ such that the columns have mean zero and unit variance.
			
			\State Calculate the spectral decomposition of the sample covariance $\mathrm{Cov}_n(\widetilde{\mathbf{Z}}) = P^\top D P$.
			\State Set the decorrelated matrix $\mathbf{Z} = \widetilde{\mathbf{Z}} P D^{-1/2}$.
			
			\State Calculate the spectral decomposition of the target empirical correlation matrix $\Sigma_r = Q^\top R Q$.
			\State The final design matrix with the exact target empirical covariance $\Sigma_r$ is given by $\mathbf{X} = \mathbf{Z} R^{1/2} Q$.
		\end{algorithmic}
		\label{algo:spurious}
	\end{algorithm}

	% --- Bibliography ---
	\bibliographystyle{plainnat} 
	\bibliography{Objective_Driven_Ensembles.bib}

\begin{thebibliography}{34}
\providecommand{\natexlab}[1]{#1}
\providecommand{\url}[1]{\texttt{#1}}
\expandafter\ifx\csname urlstyle\endcsname\relax
  \providecommand{\doi}[1]{doi: #1}\else
  \providecommand{\doi}{doi: \begingroup \urlstyle{rm}\Url}\fi

\bibitem[Bertsimas et~al.(2016)Bertsimas, King, and
  Mazumder]{bertsimas2016best}
Dimitris Bertsimas, Angela King, and Rahul Mazumder.
\newblock Best subset selection via a modern optimization lens.
\newblock \emph{The Annals of Statistics}, 44\penalty0 (2):\penalty0 813--852,
  2016.

\bibitem[Breiman(2001{\natexlab{a}})]{breiman2001random}
Leo Breiman.
\newblock Random forests.
\newblock \emph{Machine Learning}, 45\penalty0 (1):\penalty0 5--32,
  2001{\natexlab{a}}.

\bibitem[Breiman(2001{\natexlab{b}})]{breiman2001two}
Leo Breiman.
\newblock Statistical modeling: The two cultures (with comments and a rejoinder
  by the author).
\newblock \emph{Statistical science}, 16\penalty0 (3):\penalty0 199--231,
  2001{\natexlab{b}}.

\bibitem[Bühlmann and van~de Geer(2011)]{buhlmann2011statistics}
Peter Bühlmann and Sara van~de Geer.
\newblock \emph{Statistics for High-Dimensional Data: Methods, Theory and
  Applications}.
\newblock Springer Series in Statistics. Springer Science \& Business Media,
  Heidelberg, Germany, 2011.

\bibitem[Chen and Guestrin(2016)]{chen2016xgboost}
Tianqi Chen and Carlos Guestrin.
\newblock Xgboost: A scalable tree boosting system.
\newblock In \emph{Proceedings of the 22nd ACM SIGKDD International Conference
  on Knowledge Discovery and Data Mining}, pages 785--794, 2016.

\bibitem[Chipman et~al.(2010)Chipman, George, and McCulloch]{chipman2010bart}
Hugh~A Chipman, Edward~I George, and Robert~E McCulloch.
\newblock Bart: Bayesian additive regression trees.
\newblock \emph{The Annals of Applied Statistics}, 4\penalty0 (1):\penalty0
  266--298, 2010.

\bibitem[Christidis and Cohen-Freue(2026)]{R-RMSS}
Anthony Christidis and Gabriela Cohen-Freue.
\newblock \emph{RMSS: Robust Multi-Model Subset Selection}, 2026.
\newblock URL \url{https://CRAN.R-project.org/package=RMSS}.
\newblock R package version 1.2.4.

\bibitem[Christidis et~al.(2020{\natexlab{a}})Christidis, Aelst, and
  Zamar]{R-simTargetCov}
Anthony Christidis, Stefan~Van Aelst, and Ruben Zamar.
\newblock \emph{simTargetCov: Data Transformation or Simulation with Target
  Empirical Covariance Matrix}, 2020{\natexlab{a}}.
\newblock URL \url{https://r-project.org}.
\newblock R package version 1.0.1.

\bibitem[Christidis et~al.(2021)Christidis, Aelst, and Zamar]{R-splitSelect}
Anthony Christidis, Stefan~Van Aelst, and Ruben~H. Zamar.
\newblock \emph{splitSelect: Best Split Selection Modeling for Low-Dimensional
  Data}, 2021.
\newblock URL \url{https://CRAN.R-project.org/package=splitSelect}.
\newblock R package version 1.0.3.

\bibitem[Christidis et~al.(2024)Christidis, Smucler, and Zamar]{R-SplitReg}
Anthony Christidis, Ezequiel Smucler, and Ruben Zamar.
\newblock \emph{SplitReg: Split Regularized Regression}, 2024.
\newblock URL \url{https://CRAN.R-project.org/package=SplitReg}.
\newblock R package version 1.0.2.

\bibitem[Christidis et~al.(2025{\natexlab{a}})Christidis, Aelst, and
  Zamar]{R-SplitGLM}
Anthony Christidis, Stefan~Van Aelst, and Ruben Zamar.
\newblock \emph{SplitGLM: Split Generalized Linear Models}, 2025{\natexlab{a}}.
\newblock URL \url{https://CRAN.R-project.org/package=SplitGLM}.
\newblock R package version 1.0.6.

\bibitem[Christidis et~al.(2025{\natexlab{b}})Christidis, {Van Aelst}, and
  Zamar]{R-PSGD}
Anthony Christidis, Stefan {Van Aelst}, and Ruben Zamar.
\newblock \emph{PSGD: Projected Subset Gradient Descent}, 2025{\natexlab{b}}.
\newblock URL \url{https://CRAN.R-project.org/package=PSGD}.
\newblock R package version 1.0.6.

\bibitem[Christidis et~al.(2026{\natexlab{a}})Christidis, Aelst, and
  Freue]{christidis2026robust}
Anthony Christidis, Stefan~Van Aelst, and Gabriela~Cohen Freue.
\newblock Robust multi-model subset selection.
\newblock \emph{Journal of Computational and Graphical Statistics},
  2026{\natexlab{a}}.

\bibitem[Christidis(2026)]{R-srlars}
Anthony-Alexander Christidis.
\newblock \emph{srlars: Fast and Scalable Cellwise-Robust Ensemble}, 2026.
\newblock URL \url{https://CRAN.R-project.org/package=srlars}.
\newblock R package version 2.0.1.

\bibitem[Christidis et~al.(2020{\natexlab{b}})Christidis, Lakshmanan, Smucler,
  and Zamar]{christidis2020split}
Anthony-Alexander Christidis, Laks Lakshmanan, Ezequiel Smucler, and Ruben
  Zamar.
\newblock Split regularized regression.
\newblock \emph{Technometrics}, 62\penalty0 (3):\penalty0 330--338,
  2020{\natexlab{b}}.

\bibitem[Christidis et~al.(2025{\natexlab{c}})Christidis, Lakshmanan, Smucler,
  and Zamar]{christidis2025data}
Anthony-Alexander Christidis, Saptarshi Lakshmanan, Ezequiel Smucler, and Ruben
  Zamar.
\newblock Data-driven logistic regression ensembles with applications in
  genomics.
\newblock \emph{Advances in Data Analysis and Classification}, Nov
  2025{\natexlab{c}}.

\bibitem[Christidis et~al.(2025{\natexlab{d}})Christidis, Van~Aelst, and
  Zamar]{christidis2025multi}
Anthony-Alexander Christidis, Stefan Van~Aelst, and Ruben Zamar.
\newblock Multi-model subset selection.
\newblock \emph{Computational Statistics \& Data Analysis}, 203:\penalty0
  108073, 2025{\natexlab{d}}.

\bibitem[Christidis et~al.(2026{\natexlab{b}})Christidis, Pyneeandee, and
  Cohen-Freue]{christidis2026fast}
Anthony-Alexander Christidis, Jeyshinee Pyneeandee, and Gabriela Cohen-Freue.
\newblock Fast and scalable cellwise-robust ensembles for high-dimensional
  data.
\newblock \emph{arXiv preprint arXiv:2603.20940}, 2026{\natexlab{b}}.
\newblock URL \url{https://arxiv.org/abs/2603.20940}.

\bibitem[Fan et~al.(2016)Fan, Liao, and Jobson]{fan2016guarding}
Jianqing Fan, Yuan Liao, and Joanne~S. Jobson.
\newblock Guarding against spurious discoveries in high dimensions.
\newblock \emph{Journal of Machine Learning Research}, 17\penalty0
  (102):\penalty0 1--42, 2016.
\newblock URL \url{https://jmlr.org/papers/v17/16-068.html}.

\bibitem[Friedman et~al.(2010)Friedman, Hastie, and Tibshirani]{R-glmnet}
Jerome Friedman, Trevor Hastie, and Robert Tibshirani.
\newblock Regularization paths for generalized linear models via coordinate
  descent.
\newblock \emph{Journal of Statistical Software}, 33\penalty0 (1):\penalty0
  1--22, 2010.
\newblock \doi{10.18637/jss.v033.i01}.

\bibitem[Friedman(2001)]{friedman2001greedy}
Jerome~H. Friedman.
\newblock Greedy function approximation: a gradient boosting machine.
\newblock \emph{Annals of Statistics}, 29\penalty0 (5):\penalty0 1189--1232,
  2001.

\bibitem[Friedman and Popescu(2008)]{friedman2008predictive}
Jerome~H Friedman and Bogdan~E Popescu.
\newblock Predictive learning via rule ensembles.
\newblock \emph{The Annals of Applied Statistics}, 2\penalty0 (3):\penalty0
  916--954, 2008.

\bibitem[Hastie et~al.(2020)Hastie, Tibshirani, and Tibshirani]{hastie2020best}
Trevor Hastie, Robert Tibshirani, and Ryan~J Tibshirani.
\newblock Best subset, forward stepwise or lasso? analysis and recommendations
  based on extensive comparisons.
\newblock \emph{Statistical Science}, 35\penalty0 (4):\penalty0 579--592, 2020.

\bibitem[Hazimeh and Mazumder(2020)]{hazimeh2020fast}
Hussein Hazimeh and Rahul Mazumder.
\newblock Fast best subset selection: Coordinate descent and local
  combinatorial optimization algorithms.
\newblock \emph{Operations Research}, 68\penalty0 (4):\penalty0 1183--1201,
  2020.

\bibitem[Hazimeh et~al.(2022)Hazimeh, Mazumder, and Nonet]{R-L0Learn}
Hussein Hazimeh, Rahul Mazumder, and Tim Nonet.
\newblock L0learn: A scalable package for sparse learning using l0
  regularization.
\newblock \emph{Journal of Machine Learning Research}, 23\penalty0
  (266):\penalty0 1--8, 2022.
\newblock URL \url{https://jmlr.org}.

\bibitem[Song et~al.(2024)Song, Langfelder, and Horvath]{R-randomGLM}
Lin Song, Peter Langfelder, and Steve Horvath.
\newblock \emph{randomGLM: Random General Linear Model Prediction}, 2024.
\newblock URL \url{https://CRAN.R-project.org/package=randomGLM}.
\newblock R package version 1.10-1.

\bibitem[Song et~al.(2013)Song, Zhang, Qu, Luan, Ayyad, Haque,
  et~al.]{song2013random}
Peter X-K Song, Yuedong Zhang, Annie Qu, Jian Luan, Essam Ayyad, M~Emdadul
  Haque, et~al.
\newblock Random generalized linear model: a highly accurate and flexible
  ensemble predictor.
\newblock \emph{BMC bioinformatics}, 14\penalty0 (1):\penalty0 1--17, 2013.

\bibitem[Tibshirani(1996)]{tibshirani1996regression}
Robert Tibshirani.
\newblock Regression shrinkage and selection via the lasso.
\newblock \emph{Journal of the Royal Statistical Society. Series B
  (Methodological)}, 58\penalty0 (1):\penalty0 267--288, 1996.

\bibitem[Ueda and Nakano(1996)]{ueda1996generalization}
Naonori Ueda and Ryohei Nakano.
\newblock Generalization error of gradient descent learning and its stochastic
  evaluation.
\newblock \emph{IEEE Transactions on Neural Networks}, 7\penalty0 (6):\penalty0
  1482--1497, 1996.

\bibitem[Welch(1982)]{welch1982algorithmic}
W.~J. Welch.
\newblock Algorithmic complexity: three {NP}-hard problems in computational
  statistics.
\newblock \emph{Journal of Statistical Computation and Simulation}, 15\penalty0
  (1):\penalty0 17--25, 1982.

\bibitem[Wright and Ziegler(2017)]{R-ranger}
Marvin~N. Wright and Andreas Ziegler.
\newblock {ranger}: A fast implementation of random forests for high
  dimensional data in {C++} and {R}.
\newblock \emph{Journal of Statistical Software}, 77\penalty0 (1):\penalty0
  1--17, 2017.

\bibitem[Yang et~al.(2026)Yang, Van~Aelst, and Verdonck]{yang2026diverse}
Bing Yang, Stefan Van~Aelst, and Tim Verdonck.
\newblock Diverse ensemble cost-sensitive logistic regression.
\newblock \emph{European Journal of Operational Research}, 328\penalty0
  (1):\penalty0 282--294, 2026.

\bibitem[Zhao and Yu(2006)]{zhao2006model}
Peng Zhao and Bin Yu.
\newblock On model selection consistency of lasso.
\newblock \emph{Journal of Machine Learning Research}, 7\penalty0
  (89):\penalty0 2541--2567, 2006.

\bibitem[Zou and Hastie(2005)]{zou2005elasticnet}
Hui Zou and Trevor Hastie.
\newblock Regularization and variable selection via the elastic net.
\newblock \emph{Journal of the Royal Statistical Society: Series B (Statistical
  Methodology)}, 67\penalty0 (2):\penalty0 301--320, 2005.

\end{thebibliography}

\end{document}